\documentclass[twocolumn,showpacs,preprintnumbers,amsmath,amssymb]{revtex4}
\usepackage{amsmath,amssymb,graphics,epsfig,subfigure}
\usepackage{color}
\usepackage{mathrsfs}

\begin{document}
\newcommand {\nn} {\nonumber}
\renewcommand{\baselinestretch}{1.3}

\title{Universal topological classifications of black hole thermodynamics}

\author{Shao-Wen Wei$^{1,2}$ \footnote{E-mail: weishw@lzu.edu.cn},
Yu-Xiao Liu$^{1,2}$ \footnote{E-mail: liuyx@lzu.edu.cn},
Robert B. Mann$^{3}$ \footnote{E-mail: rbmann@uwaterloo.ca}}

\affiliation{$^{1}$Lanzhou Center for Theoretical Physics, Key Laboratory of Theoretical Physics of Gansu Province, and Key Laboratory of Quantum Theory and Applications of MoE, Lanzhou University, Lanzhou, Gansu 730000, People's Republic of China,\\
 $^{2}$Institute of Theoretical Physics, Research Center of Gravitation, and School of Physical Science and Technology, Lanzhou University, Lanzhou 730000, People's Republic of China,\\
 $^{3}$ Department of Physics \& Astronomy, University of Waterloo, Waterloo, Ont. Canada N2L 3G1}

\begin{abstract}
In this Letter, we investigate the universal classifications of black hole states by considering them as topological defects within the thermodynamic parameter space. Through the asymptotic behaviors of the constructed vector, our results indicate the existence of four distinct topological classifications, denoted as $W^{1-}$, $W^{0+}$, $W^{0-}$, and $W^{1+}$. Within these classifications, the innermost small black hole states are characterized as unstable, stable, unstable, and stable, respectively, while the outermost large ones exhibit an unstable, unstable, stable, and stable behavior. These classifications also display contrasting thermodynamic properties in both low and high Hawking temperature limits. Furthermore, we establish a systematic ordering of the local thermodynamically stable and unstable black hole states as the horizon radius increases for a specific topological classification. These results reveal the universal topological classifications governing black hole thermodynamics, providing valuable insights into the fundamental nature of quantum gravity.
\end{abstract}

\keywords{Classical black hole, thermodynamics, topology.}

\pacs{04.70.Dy, 04.70.Bw, 05.70.Ce}

\maketitle

{\emph{Introduction}.}---Black hole thermodynamics remains a highly regarded and actively researched topic, drawing significant attention ever since its establishment. Of particular interest are the underlying phase transitions in anti-de Sitter (AdS) spacetime, notably the Hawking-Page phase transition \cite{HawkingPage}, which has crucial implications for gauge theory via the AdS/CFT correspondence. Exploration of extended phase space thermodynamics \cite{Kubiznak:2016qmn} subsequently attracted considerable attention, and led to the discovery of numerous fascinating phase transitions and phase structures \cite{Kastor, Kubiznaka,Weicoexistence}. These discoveries have recently inspired further inquiries into various aspects of the holographic dual of gravitational thermodynamics, including the Euler relation, the role of the central charge, and its higher dimensional origin \cite{Cong,Gao,Frassino,Ahmed,Gong,Kubiznak}.

Despite such significant progress, uncovering the universal properties of black hole thermodynamics remains a challenge. Recently it has become appreciated that topology provides a promising approach to this end \cite{Weicr,WeiDefect}: black hole solutions can be regarded as topological defects within the thermodynamic parameter space. Each black hole state is characterized by a winding number that quantifies its topological property, which appeared to be one of three types.

Unraveling the complete universal topological classification of black holes and their associated thermodynamic properties are therefore worth considering. Such a classification scheme holds significant potential for enhancing our understanding of black hole thermodynamics and for providing insights into the fundamental nature of quantum gravity. As we shall see, our findings suggest the existence of four distinct classes, challenging the conventional notion of only three topological classifications. We note that topological analysis has proven useful in constructing light rings for ultra-compact objects and stationary black holes, revealing that there exists at least one unstable light ring for the Kerr-like black hole \cite{Cunhaa,Cunhab}.

The topology of black hole thermodynamics primarily relies on the generalized off-shell free energy proposed by York \cite{York}. Within this framework, a black hole, characterized by mass $M$ and entropy $S$, is placed in a cavity with temperature $1/\tau$. The generalized free energy is given by $\mathcal{F} = M - S/\tau$; this can also be obtained by evaluating the gravitational path integral \cite{Lir}. Only when $\tau = \beta = 1/T$, where $T$ represents the Hawking temperature of the black hole, does the free energy become an on-shell quantity. By incorporating an additional parameter $\Theta \in (0, \pi)$, we can define a two-component vector $\phi$ in terms of the gradient of $\mathcal{\tilde{F}} = \mathcal{F} + 1/\sin\Theta$,
\begin{eqnarray}
 \phi=\left(\frac{\partial \mathcal{\tilde{F}}}{\partial r_{h}}, \frac{\partial \mathcal{\tilde{F}}}{\partial\Theta}\right).
\end{eqnarray}
A detailed study reveals that the black hole states precisely reside at the zero points (or defects) of the vector $\phi$. Subsequently, employing Duan's $\phi$-mapping topological current theory \cite{Duana, Duanb}, a topological charge, known as the winding number $w$, can be assigned to each zero point or black hole state \cite{WeiDefect}.

Using this framework the positive or negative heat capacity of a black hole state can be shown to respectively correspond to a positive or negative winding number, $w = +1$ and $w = -1$, associated with locally stable and unstable black hole states. Summing all these winding numbers yields the topological number
\begin{eqnarray}
 W=\sum_{i=1}^{N}w_{i},
\end{eqnarray}
for a given black hole system,
where $w_{i}$ represents the winding number associated with the $i$-th zero point of $\phi$, with a total of $N$ zero points. The topological number $W$ characterizes each black hole system and so can be used as a classification parameter.

{\emph{Topological classifications}.}---The property of the defect curve, the collection of defects plotted in the $(r_h,\tau)$ plane, plays an important role in calculating the topological classification. Consider for simplicity the case with a single defect curve. Quantitative investigations have demonstrated that the topological classifications are determined by the asymptotic behaviour of the Hawking temperature for both small and large black holes along this curve \cite{WeiDefect}, namely the limits
\begin{eqnarray}
 r_{h}\rightarrow r_{m},\quad \infty
\end{eqnarray}
where $r_{m}$ is the minimal radius of the black hole horizon, which may or may not vanish. For example a Reissner-Nordstr\"{o}m (RN) black hole with fixed charge $Q$ will have $r_{m}=M=Q=r_e$ (the extremal case), whereas $r_{m} = 0$ for a Schwarzschild black hole.

In order to illustrate the asymptotic limits of the Hawking temperature, let us consider the following line element
\begin{eqnarray}
 ds^2=-f(r)dt^2+f^{-1}(r)dr^2+r^2 d\Omega^{2},
\end{eqnarray}
for an example. In general, we assume $f=1-2M/r+h(r)$. Taking $h(r)=0$, the Schwarzschild black hole will be recovered, and the Minkowski space time will be further recovered if $M=0$. The black hole horizon is located at the largest root of $f(r_{h})=0$ giving $M=r_{h}(1+h(r_{h}))/2$. The Hawking temperature is
\begin{eqnarray}
 T=\frac{ f'(r_{h})}{4\pi}=\frac{1}{4\pi r_{h}}\left(1+h(r_{h})+r_{h}h'(r_{h})\right),
\end{eqnarray}
where the prime denotes differentiation with respect to $r$. In general, we see that the Hawking temperature will either vanish or diverge in the limits $r\to\infty$ and $r\to 0$ unless $h(r_{h})$ is fine-tuned to cancel the contribution of $r_{h}$ in the denominator. Moreover, for a degenerate horizon $T(r_{m})=0$. Therefore at these asymptotic limits black holes exhibit either vanishing or infinite temperature, yielding four possibilities, as follows:
\begin{eqnarray}
 &case\; I\;:& \beta(r_{m})=0,\quad\;\; \beta(\infty)=\infty,\label{a1}\\
 &case\; II:& \beta(r_{m})=\infty,\quad \beta(\infty)=\infty,\\
 &case\; III:& \beta(r_{m})=\infty,\quad \beta(\infty)=0,\\
  &case\; IV:& \beta(r_{m})=0,\quad\;\; \beta(\infty)=0 \label{a4}
\end{eqnarray}
for the asymptotic behaviour of the inverse temperature $\beta(r_h)$. The defect curve $\beta(r_{h})$ is required to be analytic in the range ($r_m$, $\infty$).

We now consider the asymptotic behaviour of $\phi$ at the boundary corresponding to (\ref{a1})-(\ref{a4}). This boundary can be described via the contour $C=I_1\cup I_2\cup I_3\cup I_4$, where $I_{1}=\{r_{h}=\infty,\;\Theta\in(0, \pi)\}$, $I_{2}=\{r_{h}=(\infty,\;r_{m}),\;\Theta=\pi\}$, $I_{3}=\{r_{h}=r_{m},\;\Theta\in(\pi, 0)\}$, and $I_{4}=\{r_{h}=(r_{m},\;\infty),\;\Theta=0\}$, which encompasses all possible parameter regions.

The construction of $\phi$ implies that its direction is orthogonal to $I_2$ and $I_4$ \cite{Cunhab, WeiDefect}, and so the asymptotic behaviour of interest is along $I_1$ and $I_3$. The $r_{h}$ component is
\begin{eqnarray}
 \phi^{r_{h}}=\frac{\partial\tilde{\mathcal{F}}}{\partial r_{h}}
     = \frac{\partial M}{\partial S}\frac{\partial S}{\partial r_{h}}-\frac{1}{\tau}\frac{\partial S}{\partial r_{h}}=\frac{\partial S}{\partial r_{h}}\left(\frac{1}{\beta}-\frac{1}{\tau}\right),
\end{eqnarray}
using the first law of black hole thermodynamics. Since $\tau$ represents the temperature of the cavity it is a fixed finite positive constant. Consequently the asymptotic behavior of $\phi^{r_{h}}$ is entirely dependent on $\beta$ for positive $\frac{\partial S}{\partial r_{h}}$. As $\beta$ approaches $0$, $\phi^{r_{h}}$ must become positive and as $\beta$ approaches $\infty$, $\phi^{r_{h}}$ must become negative. Consequently, as $r_{h}$ approaches $r_{m}$ and $\infty$, the direction of $\phi$ is either right- or leftward, with some inclination depending on the value of $\phi^{\Theta}$.

\begin{table}[]
\setlength{\tabcolsep}{2.5mm}{\begin{tabular}{cccccc}\hline\hline
     & $I_{1}$ &$I_{2}$ & $I_{3}$ & $I_{4}$ &$W$ \\\hline\hline
 case I      & $\leftarrow$ & $\uparrow$& $\rightarrow$ & $\downarrow$&-1 \\
 case II   & $\leftarrow$   &$\uparrow$ & $\leftarrow$     &$\downarrow$ &0 \\
 case IIII      &  $\rightarrow$   & $\uparrow$& $\leftarrow$ & $\downarrow$&+1\\
 case IV   & $\rightarrow$ &$\uparrow$ & $\rightarrow$ & $\downarrow$&0 \\\hline\hline
\end{tabular}
\caption{The directions indicated by the arrows of $\phi^{r_{h}}$ are shown for the four segments. The corresponding topological number for each case is also listed in the last column.}\label{tab}}
\end{table}

 In Table I the four possible direction pairs on the segments $I_1$ and $I_3$ are listed, along with the topological number $W$ for each case. This evidently reveals three distinct topological classifications, apparently confirming the conjecture proposed in \cite{WeiDefect} and commented on in \cite{Liu}.

The Schwarzschild black hole serves as a representative example for case I. For a given $\tau$, there exists only one black hole state characterized by a negative heat capacity. This state possesses a local winding number of -1, in perfect agreement with its total topological number $W=-1$.

For case II the RN black hole serves as a representative example. Here the presence of electric charge introduces significant modifications, since small charged black holes have positive heat capacity, whereas large ones retain negative heat capacity. A degenerate point occurs at a certain $\beta_*$ \cite{WeiDefect}. Below this point, no black hole states exist, resulting in a trivial topology. For large values of $\beta$ two black hole states, small and large, emerge, which are thermodynamically stable and unstable respectively. Consequently they possess opposite winding numbers, resulting in a vanishing topological number. This observation emphasizes the remarkable fact that the temperature of the cavity serves as an external parameter and does not influence the intrinsic properties of the black hole.

In the third case, the topological number is +1, indicating the presence of an additional locally stable black hole state for a fixed $\tau$, an example being the RN-AdS black hole. The presence of a negative cosmological constant introduces another stable state in addition to the two present in the RN black hole, resulting in the expected topological number $W=+1$. When the cosmological constant is small, we can have one unstable and two stable black hole states for moderate values of $\tau$. Conversely, if the cosmological constant is sufficiently large, the unstable black hole state is entirely excluded. In such cases, only one stable black hole state remains for arbitrary $\tau$. The topological number always remains as +1, independent of both $\tau$ and the cosmological constant.

\begin{figure}
\center{\subfigure[The RN black hole.]{\label{RN}\includegraphics[width=4cm]{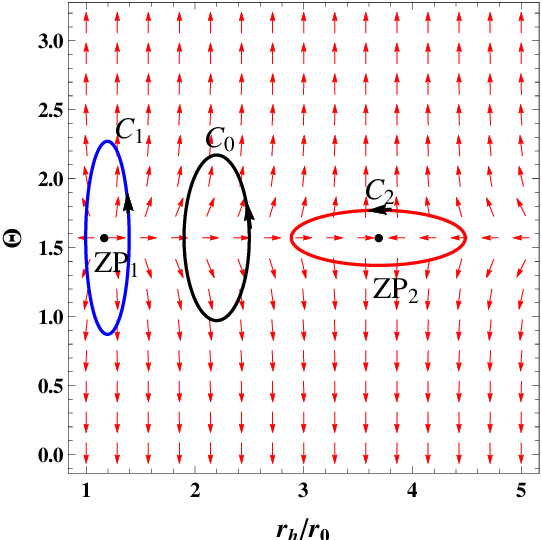}}
\subfigure[The Schwarzschild-AdS black hole.]{\label{SchAdS}\includegraphics[width=4cm]{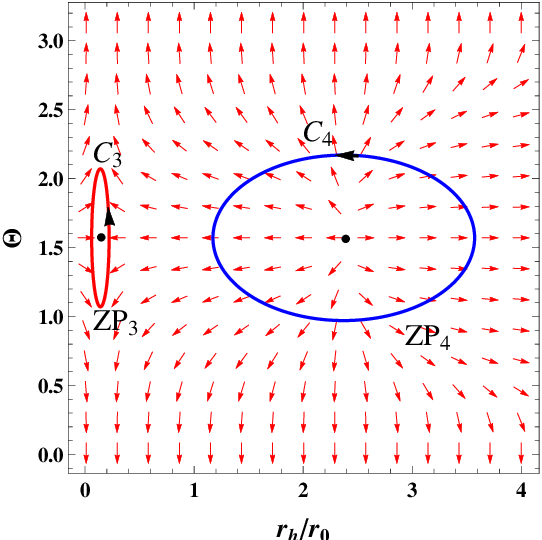}}}
\caption{The unit vectors on a portion of the $r_{\text{h}}$-$\Theta$ plane. The red arrows represent the direction of $\phi$. The zero points (ZPs) are marked with black dots and all contours are counterclockwise by convention. The blue contours $C_i$ ($i$=1-4) are closed loops enclosing each zero point, while $C_{0}$ does not enclose any zero point. (a) The unit vector field for the RN black hole with $\tau/r_0=50$ and charge $Q$=1. (b) The unit vector field for the Schwarzschild-AdS black hole with $\tau/r_0=1.67$ and the AdS radius $l/r_0=1$.
}\label{pSchAdS}
\end{figure}

\begin{figure}
\center{\subfigure[The RN black hole.]{\label{VecRN}\includegraphics[width=4cm]{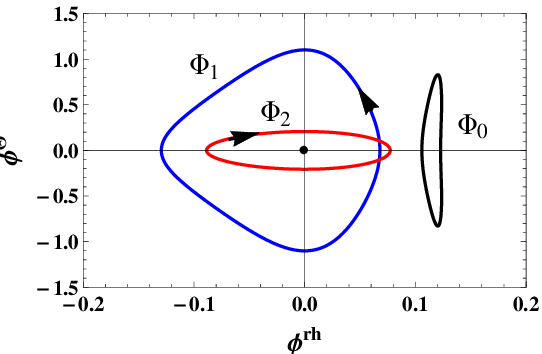}}
\subfigure[The Schwarzschild-AdS black hole.]{\label{VecSchAdS}\includegraphics[width=4cm]{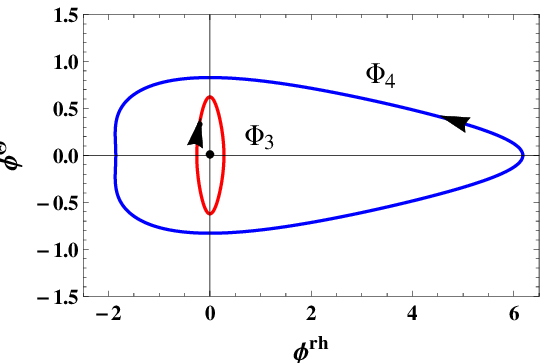}}}
\caption{Contours $\Phi_i$ that map the change in the components of $\phi$ as the contours $C_{i}$ ($i$=0-4) in figure~\ref{pSchAdS} are traversed, where (a) is the RN black hole and (b) is the Schwarzschild-AdS black hole, and the origin corresponds to the zero points of $\phi$. The black arrows indicate the sense of rotation that the vector $\phi$ undergoes as it traverses a contour in Fig.~\ref{pSchAdS}. Along $C_0$, $\phi$ has no net rotation; this contour has no arrow and does not enclose the origin, so the corresponding winding number is zero. Along contours $C_1$ and $C_4$, the changes in the components of $\phi$ force the closed loops $\Phi_1$ and $\Phi_4$ to be counterclockwise, and thus the winding number is +1. For $C_2$ and $C_3$, the behaviours are reversed, and the winding number is -1. }\label{pVecSchAdS}
\end{figure}

Table \ref{tab} indicates that case IV also has a vanishing topological number, the same as case II. However, these cases display contrasting behaviour at small and large $r_{h}$. If $W=0$ stable and unstable black hole states appear in pairs (if they exist). However the order in which these states occur differs between the second and fourth cases. An example in the case IV class is the Schwarzschild-AdS (SAdS) black hole, whose unit vector in the $\Theta$-$r_{h}$ plane we present in Fig.~\ref{pSchAdS}, along with that of the RN black hole for comparison. Both black holes obviously exhibit two zeros of $\phi$. However the small RN black hole has positive winding number whereas the small SAdS black hole has negative winding number. These signs are each reversed for the respective large black holes (at the right of each figure).

Consider next the changes in the components ($\phi^{r_{h}},\phi^{\Theta})$ of $\phi$ along each of the contours in Fig.~\ref{pSchAdS}, shown in Fig.~\ref{pVecSchAdS}. The zero points of $\phi$ are at the origin, and mapping these changes along each $C_{i}$ yields a closed contour $\Phi_i$ in this vector space. The direction of a contour in Fig.~\ref{pVecSchAdS} corresponds to sense of rotation of $\phi$ as its corresponding (counterclockwise) contour in Fig.~\ref{pSchAdS} is traversed. Contours with positive winding numbers in Fig.~\ref{pSchAdS} are mapped to counterclockwise contours in Fig.~\ref{pVecSchAdS}, whereas those with negative winding numbers are mapped to clockwise contours. The vector $\phi$ undergoes no net rotation about curves in Fig.~\ref{pSchAdS} that do not surround a zero point; such curves have no direction and do not surround the origin in Fig.~\ref{pVecSchAdS}. For the RN black hole, the winding numbers of the first and second zeros are +1 and -1, respectively; for the Schwarzschild-AdS black hole, the first and second zeros possess winding numbers -1 and +1, respectively.

This clear distinction allows us to employ order as a means to further classify these two cases, where class II can be denoted as $W^{0+}$ and class IV as $W^{0-}$. The signs $+$ or $-$ respectively indicate whether the small black hole state has a winding number of +1 or -1.

In summary, we now have four topological classifications of black hole thermodynamics:
\begin{eqnarray}
 W^{1-}, \quad W^{0+}, \quad W^{0-}, \quad W^{1+}.
\end{eqnarray}
It is noteworthy that while the second and third classifications, respectively exemplified by RN and SAdS black holes, share the same topological number, their thermodynamic characteristics vary significantly at small and large $r_{h}$. Consequently, our classifications offer an effective means of discerning between them.

{\emph{Black hole state systematic orderings}.}--- Following the classification, a fundamental question arises: are there universal thermodynamic patterns associated with each topological classification? To address this issue, we turn to analyze the systematic ordering within these topological classifications.

In case I, there is at least one black hole state with negative heat capacity and a winding number of $-1$. If any additional black hole states exist, they must appear in pairs. Since the signs of the heat capacities alternate with increasing $r_{h}$, the smallest and largest states both correspond to unstable black holes, and the winding numbers associated with the different zero points follow (from smallest to largest values of $r_h$) the pattern [-, (+, -), ..., (+, -)], where the ellipsis denotes pairs of (+, -) winding numbers. In case III this behaviour is reversed, and the order is [+, (-, +), ..., (-, +)]; the innermost and outermost states both are the stable black holes. The second and fourth cases exhibit winding number orders of [+, (-, +), ..., -] and [-, (+, -), ..., +] respectively. For simplicity, these four cases can be labeled based on the signs of the innermost and outermost winding numbers:
\begin{eqnarray}
 W^{1-}=[-, -], \quad W^{0+}=[+, -], \nonumber\\
 W^{0-}=[-, +],  \quad  W^{1+}=[+, +].\label{aidos}
\end{eqnarray}

{\it Universal thermodynamic behaviour}---The topological classification \eqref{aidos} furnishes a universal description of the behaviour of black hole states in the limits of small and large black holes, as well as low and high temperatures.

Recall that the inner- and outermost states respectively correspond to the limits of small and large black holes. The classification (\ref{aidos}) indicates that the innermost black hole states are characterized as unstable, stable, unstable, and stable for the respective $W^{1-}$, $W^{0+}$, $W^{0-}$, and $W^{1+}$ classes. Likewise, the outermost black hole states are respectively unstable, unstable, stable, and stable.

Concerning the low temperature limit $\beta\to\infty$, the $W^{1-}$ class has a large black hole that is thermodynamically unstable due to its topological number. The $W^{1+}$ class is characterized by a single stable small black hole state. The $W^{0+}$ class has one unstable large black hole state and one stable small black hole state, whereas there is no black hole state in the $W^{0-}$ class at the low temperature limit.

At the high temperature $\beta\to 0$ limit, it is now the $W^{0+}$ class that does not possess any black hole state, whereas the $W^{0-}$ class displays both an unstable small black hole state and a stable large black hole state. The $W^{1-}$ class exhibits an unstable small black hole state, and the $W^{1+}$ class features a stable large black hole state.

{\emph{Degenerate points}.}--- The black hole states correspond to the zero points of the vector $\phi$. A zero point for which $\frac{\partial\beta}{\partial r_{h}}=0$ is degenerate, and corresponds to either a generation or annihilation point \cite{Fu}. Each can be assigned a vanishing winding number. Degenerate points are expected for the $W^{0+}$ and $W^{0-}$ classes, which exhibit pairs of black hole states. The RN black hole system displays a generation point, whereas the Schwarzschild-AdS black hole system has an annihilation point.

However the number of pairs of black hole states is not directly determined by the topological number and so can be arbitrary. Since the defect curve is continuous and differentiable, annihilation and generation points are expected to appear in pairs for the classifications $W^{1-}$ and $W^{1+}$. By contrast, the topological classifications $W^{0+}$ and $W^{0-}$ respectively exhibit one additional generation or annihilation point.

Since
\begin{eqnarray}
 \frac{\partial\beta}{\partial r_{h}}=-\frac{1}{T^2}\frac{\partial S}{\partial r_{h}}\left(\frac{\partial T}{\partial S}\right)=-\frac{1}{T}\frac{\partial S}{\partial r_{h}}\frac{1}{\mathcal{C}}
\end{eqnarray}
a degenerate point coincides with the Davies point \cite{Davies}, which is a first-order phase transition where the heat capacity $\mathcal{C}$ becomes divergent.
\begin{widetext}
\begin{center}
\begin{table}
\begin{tabular}{cccccc}\hline\hline
     & innermost & outermost & low $T$ & high $T$ & DG \\\hline\hline
    $W^{1-}$     & unstable & unstable & unstable large & unstable small  & in pairs \\
    $W^{0+}$    & stable     & unstable & unstable large+stable small    & no & one more GP \\
    $W^{1+}$    & stable     & stable     & stable small  & stable large & in pairs\\
    $W^{0-}$     & unstable & stable     & no & unstable small+stable large  & one more AP\\\hline\hline
\end{tabular}
\caption{Thermodynamical properties of these black hole states for different topological classifications. The last column represents the property of the degenerate points. DG, AP, and GP are for degenerate point, annihilation point, and generation point, respectively.}\label{tabb}
\end{table}
\end{center}
\end{widetext}

{\emph{Conclusions}.}--- We have demonstrated the existence of four distinct topological classifications: $W^{1-}$, $W^{0+}$, $W^{0-}$, and $W^{1+}$ in topological black hole thermodynamics. Any given black hole system satisfying one of the asymptotic behaviours (\ref{a1})-(\ref{a4}) must be in one of these mutually exclusive classes. A crucial point is that although $W^{0+}$ and $W^{0-}$ share the same topological number, they are distinct classes. For each topological classification, we obtained its thermodynamic behaviours in the small and large black hole limits, as well as at the low and high temperature limits. We provide a concise overview of the results in Table II. This classification scheme provides valuable insights into understanding the universality and nature of black hole thermodynamics. Moreover, it serves as a screening tool for selecting black hole systems satisfying the demands of quantum gravity or other gravitational theories.

If the asymptotic behaviors (\ref{a1})-(\ref{a4}) are not satisfied by a given black hole system, topological phase transitions can occur. A notable example occurs for multi-charged AdS black holes in gauged supergravities, where finite temperature can be attained at the small black hole limit \cite{Wu}. Consequently, topological phase transitions occur at specific temperatures within these systems.

Alternatively, even if the asymptotic behaviors (\ref{a1})-(\ref{a4}) hold, there may not be a single defect curve, in which case certain modifications occur. In particular, at the end points of these separated ranges, $\beta$ either vanishes or diverges, and two defect curves either meet at an annihilation point or emerge from a generation point; multiple defect curves can thus occur, as exemplified in Ref. \cite{Chen}. In such scenarios, more than one black hole state can survive at the limits of small or high temperatures. Nevertheless, the total topological number remains unchanged, indicating that any additional black hole states must appear in pairs. Furthermore, the number of degenerate points is also subject to change, and definitive conclusions regarding their specific quantities cannot be drawn.

This topological perspective we have presented offers a profound understanding of the universal properties of black hole thermodynamics. It also highlights the distinctive features that enable certain topological classifications to satisfy the requirements imposed by potential astronomical observations.

{\emph{Acknowledgements}.}---This work was supported by the National Natural Science Foundation of China (Grants No. 12075103, No. 12475056, and No. 12247101), the 111 Project (Grant No. B20063), and by the Natural Sciences and Engineering Research Council of Canada.

\end{document}